\documentclass[aps,prd,preprint,groupedaddress]{revtex4-1}

 \usepackage{graphicx}
 \usepackage{epstopdf}
 \usepackage{color}
 \usepackage[colorlinks=true, urlcolor=navyblue, linkcolor=navyblue, citecolor=navyblue]{hyperref}
\usepackage{lineno}

\def\Dslash{\raise.15ex\hbox{/}\kern-.7em D}
\def\Pslash{\raise.15ex\hbox{/}\kern-.7em P}

\newcommand{\beq}{\begin{equation}}
\newcommand{\enq}{\end{equation}}
\newcommand{\beqa}{\begin{eqnarray}}
\newcommand{\beqast}{\begin{eqnarray*}}
\newcommand{\enqa}{\end{eqnarray}}
\newcommand{\enqast}{\end{eqnarray*}}
\newcommand{\nn}{\nonumber}
\newcommand{\req}[1]{(\ref{#1})}

\renewcommand{\arraystretch}{1.3}

\newcommand{\mbf}[1]{\mathbf{#1}}

\newcommand{\half}{{\frac{1}{2}}}

\newcommand{\bec}{\begin{center}}
\newcommand{\enc}{\end{center}}
\newcommand{\beqo}{\begin{quote}}
\newcommand{\enqo}{\end{quote}}

\newcommand{\al}{\alpha}
\newcommand{\be}{\beta}

\newcommand{\de}{\delta}

\newcommand{\ze}{\zeta}

\newcommand{\la}{\lambda}

\newcommand{\si}{\sigma}

\newcommand{\ph}{\phi}

\newcommand{\vp}{\varphi}

\newcommand{\De}{\Delta}

\newcommand{\La}{\Lambda}

\newcommand{\Si}{\Sigma}

\newcommand{\thalf}{\textstyle \frac{3}{2}}
\newcommand{\fhalf}{\textstyle \frac{5}{2}}

\definecolor{navyblue}{rgb}{0,0.08,0.45}
\definecolor{darkred}{rgb}{0.7,0.0,0.0}
\definecolor{darkgreen}{rgb}{0,0.6,0.2}

\begin{document}


\preprint{SLAC--PUB--16882}

\title{Supersymmetry Across the Light and \\Heavy-Light Hadronic Spectrum II}

\author{Hans G\"unter Dosch}
\affiliation{Institut f\"ur Theoretische Physik, Philosophenweg
16, 69120 Heidelberg, Germany}
\email[]{h.g.dosch@thphys.uni-heidelberg.de}

\author{Guy F.  de T\'eramond}
\affiliation{Universidad de Costa Rica, 11501 San Pedro de Montes
de Oca, Costa Rica} \email[]{gdt@asterix.crnet.cr}

\author{Stanley J. Brodsky}
\affiliation{SLAC National Accelerator Laboratory, Stanford
University, Stanford, California 94309, USA}
\email{sjbth@slac.stanford.edu}

\date{\today}

\begin{abstract}

We extend our analysis of the implications of hadronic supersymmetry for heavy-light hadrons in light-front holographic QCD. Although conformal symmetry is strongly broken by the heavy quark mass, supersymmetry and the holographic embedding of semiclassical light-front dynamics derived from five-dimensional anti-de Sitter (AdS) space nevertheless determines the form of the confining potential  in the light-front Hamiltonian to be harmonic. The resulting light-front bound-state equations  lead to a heavy-light Regge-like spectrum for both mesons and baryons. The confinement hadron mass scale and their Regge slopes depend, however, on the mass of the heavy quark in the meson or baryon as expected from Heavy Quark Effective Theory (HQET). This procedure reproduces the observed spectra of heavy-light hadrons with good precision and makes predictions for yet unobserved states.

\end{abstract}

 \maketitle

\tableofcontents

\section{Introduction \label{intro}}

In a series of recent articles~\cite{deTeramond:2014asa, Dosch:2015nwa, Dosch:2015bca, Brodsky:2016yod}, we have shown that superconformal algebra allows  the construction of  relativistic light-front  (LF) semiclassical bound-state equations in physical spacetime which can be embedded in a higher dimensional classical gravitational theory. This new approach to hadron physics incorporates basic nonperturbative properties which are not apparent from the chiral QCD Lagrangian; it includes the emergence of a mass scale and confinement out of a classically scale-invariant theory, the occurrence of a zero-mass bound state, universal Regge trajectories for both mesons and baryons,  and the breaking of chiral symmetry in the hadron spectrum. This holographic approach to hadronic physics gives remarkable connections between the light meson and nucleon spectra~\cite{Dosch:2015nwa}, as well as specific relations which can be derived for heavy-light hadrons. Remarkably,
even though heavy quark masses break conformal invariance, an underlying dynamical supersymmetry still holds~\cite{Dosch:2015bca}.

Our analysis is based on a procedure developed by de Alfaro, Fubini and Furlan, and Fubini and Rabinovici~\cite{deAlfaro:1976je, Fubini:1984hf, Brodsky:2013ar,deTeramond:2014asa, Dosch:2015nwa}.  In our approach, it leads to the  natural emergence of a mass scale into the Hamiltonian of a theory while retaining essential elements of both conformal invariance and supersymmetry.   In the case of  superconformal (graded) algebra, a generalized Hamiltonian can be constructed as a linear superposition of superconformal generators which carry different dimensions; the Hamiltonian thus remains within the superconformal algebraic structure.  This procedure determines  a unique form of a quark confinement potential in the light-front Hamiltonian for light mesons and baryons, and it  reproduces quite well significant features of the hadron spectrum and dynamics. The resulting bound-state equations depend explicitly on orbital angular momentum,  and thus chiral symmetry is broken from the outset in the Regge excitation spectra: The $\rho$ meson and the nucleon have no chiral partners. A striking feature of the formalism is that the supermultiplets consist of a meson wave function with internal LF angular momentum $L_M$ and a corresponding baryon wave function with angular momentum $L_B = L_M - 1$  and identical mass. The lightest meson state with $ L_M=0$ and total quark spin zero is massless in the chiral limit and is identified with the pion;  it has no supersymmetric partner.

It is not known why the effective theory based on superconformal quantum mechanics and its light-front holographic embedding captures so well essential aspects of the confinement dynamics of QCD. However, underlying aspects of the superconformal holographic construction, conformal symmetry and supersymmetry, as well as the LF cluster decomposition required by the holographic embedding, could help us understand fundamental features of  QCD in its nonperturbative domain.

As it is the case for conformal quantum mechanics~\cite{deAlfaro:1976je}, where the action remains invariant under conformal transformations, classical QCD in the limit of massless quarks has no mass scale, but confinement and a mass gap can emerge from its quantum embodiment. The cluster decomposition of the constituents of baryons corresponding to a quark-diquark structure is necessary in order to describe baryons in light-front holographic QCD (LFHQCD) since there is only a single holographic variable~\cite{Brodsky:2014yha}. The required LF clustering follows from the mapping of anti-de Sitter (AdS) equations to QCD bound-state equations in light-front physics~\cite{Brodsky:2006uqa}, where one identifies the holographic variable $z$ in the AdS classical gravity theory with the boost-invariant transverse separation $\zeta$ between constituents in the light-front quantization scheme~\cite{Dirac:1949cp, Brodsky:1997de}. In the case of mesons,  $\zeta^2 = b^2_\perp x(1-x)$ is conjugate to the invariant mass of the  $q \bar q$ in the LF wave function; it is the invariant variable of the LF Hamiltonian theory~\cite{notecluster}.  The resulting symmetry between mesons and baryons is consistent with an essential feature of color $SU(N_C)$:  a cluster of $N_C-1$ constituents can be in the same color representation as the anti-constituent; for $SU(3)$ this means $\bf \bar 3 \in  \bf 3 \times \bf 3$ and $\bf  3 \in  \bf \bar3 \times \bf \bar3$. Thus, emerging hadronic supersymmetry can be rooted in the dynamics of color $SU(3)$~\cite{deTeramond:2016bre, notesusy1}.

Our basic model describes  the confinement of massless quarks \cite{deTeramond:2014asa,Dosch:2015nwa,Brodsky:2016yod}. Indeed, for light quark masses it makes sense to apply superconformal dynamics and to treat the quark masses as perturbations: The dynamics is then not significantly changed for nonzero  quark mass, and the resulting confinement scale remains universal for the resulting 
hadronic bound states~\cite{Brodsky:2016yod}. In contrast, in the case of heavy quark masses, we cannot rely on  conclusions drawn from conformal symmetry; however, the presence of a heavy mass 
need not  also break  supersymmetry since it can stem from the dynamics of color confinement~\cite{notesusy2}. Indeed, as we have shown in Ref.~\cite{Dosch:2015bca},  supersymmetric relations between the meson and baryon masses still hold  to a good approximation even for heavy-light, {\it i.e.},  charm and bottom, hadrons.

In addition to the constraints imposed by supersymmetry, we will use additional features imposed by the holographic embedding in order to constrain the specific form of the confinement potential in the heavy-light sector.  We will also  use the heavy-quark flavor symmetry of QCD~\cite{Isgur:1991wq} to determine the dependence of the confinement scale on the heavy quark mass in the heavy mass limit,  since this symmetry is compatible with the light-front holographic approach~\cite{Branz:2010ub}.  Other holographic approaches to the heavy-light sector,  including the recent holographic approach  given in  Ref.~\cite{Liu:2016iqo}, which includes chiral  and heavy quark symmetry, have been been proposed in Refs.~\cite{Paredes:2004is, Erdmenger:2006bg, Erdmenger:2007vj, Herzog:2008bp, Bai:2013rza, Ahmady:2015yea}.

Light quark masses are not only essential for approximate conformal symmetry, but they also  guarantee the decoupling of transverse degrees of freedom -- expressed through the LF variable $\ze$ in the hadron LF wave function -- from the  longitudinal degrees of freedom which depends on the longitudinal LF momentum fraction $x$~\cite{deTeramond:2008ht}. The holographic mapping derived from the geometry of AdS space encodes the kinematics in 3+1 physical spacetime, and the modification of the AdS action -- usually described for mesons in terms of a dilaton profile $\varphi(z)$ --  generates the confining LF potential $U(z)$ in the light-front bound-state equations~\cite{deTeramond:2013it}.

Since light  constituents are present in the heavy-light bound states of mesons or baryons, the system is still ultrarelativistic;  thus the heavy-light bound states need to be described by relativistic LF bound-state equations. This means that the heavy-light system has properties common to both  the chiral  and the heavy-quark flavor sectors~\cite{Isgur:1991wq, Liu:2016iqo}.  It also suggests that we can holographically connect the supersymmetric theory to a modified AdS space; this  will be possible if the separation of the dynamical and kinematical variables also persist, at least to a good approximation, in the heavy-light domain. As we will show, we can again derive a unique confinement potential for both mesons and baryons in the heavy-light sector, even when conformal symmetry is broken by a heavy quark mass.  The resulting embedding  leads to a LF harmonic confinement potential for the heavy-light hadrons and thus  to Regge trajectories; however, as we shall show,  the confinement scale  and Regge slope depends on the mass of the heavy quark.   We will investigate this  dependence  using Heavy Quark Effective Theory (HQET)~\cite{Isgur:1991wq}. The procedure
discussed in this article not only reproduces the observed data to a reasonable accuracy, but it also allows us to make predictions for yet unobserved states.

This article is organized as follows: In Sec.~\ref{SLFH} we will briefly review the construction of the  LF Hamiltonian from supersymmetric quantum mechanics~\cite{Witten:1981nf} using the methods developed in Refs.~\cite{Fubini:1984hf,deTeramond:2014asa, Dosch:2015nwa}. In Sec.~\ref{HLS}  we extend  our approach to systems containing a heavy, charm or bottom, quark. Notably, we  discuss the constraints imposed by the holographic embedding on the supersymmetric potential, which in turn determine the form of the light front potential.  We compare  our predictions with experiment in 
Sec.~\ref{exp}, and in Sec.~\ref{sechqet} we discuss the constraints on the confinement scale imposed by HQET. Some final comments are given in Sec.~\ref{sc}. In the Appendix~\ref{app} we give expressions for the LF wave functions and hadron distribution amplitudes which are compatible with our general approach. This article is the continuation of Ref.~\cite{Dosch:2015bca}.

\section{The Supersymmetric Light-Front Hamiltonian \label{SLFH}}

The  light-front  Hamiltonian derived in the framework of
supersymmetric quantum mechanics \cite{Witten:1981nf,
Cooper:1994eh}  contains two fermionic generators, the
supercharges, $Q$ and $Q^\dagger$ with the anticommutation
relations \beq \{Q,Q\} = \{Q^\dagger,Q^\dagger\}=0, \enq and the
Hamiltonian $H$ \beq   \label{H} H=  \{Q,Q^\dagger\}, \enq which
commutes with the fermionic generators ${[Q, H]}  = [Q^\dagger, H]
= 0$, closing the graded Lie algebra. Since the Hamiltonian $H$
commutes with $Q^\dagger$, it  follows that the states  $\vert
\phi \rangle$ and $Q^\dagger  \vert \phi \rangle$ have identical
non-vanishing eigenvalues. In addition, if $|\ph_0 \rangle$ is an
eigenstate of $Q$ with zero eigenvalue, it is annihilated by the
operator $Q^\dagger$: $Q^\dagger|\ph_0 \rangle = 0$. This implies
that the lowest mesonic state on a given trajectory has no
supersymmetric baryon partner \cite{Dosch:2015nwa}. This shows the
special role of the pion in the supersymmetric approach to
hadronic physics as a unique state of zero mass in the chiral
limit.

In matrix notation \beq Q = \left(\begin{array}{cc}
0 & q\\
0 & 0\\
\end{array}
\right) , \quad  \quad Q^\dagger=\left(\begin{array}{cc}
0 & 0\\
q^\dagger & 0\\
\end{array} \right),
\enq and \beq H= \left(\begin{array}{cc}
q \, q^\dagger &  0\\
0 & q^\dagger q \\
\end{array}
\right) , \enq with \beqa \label{qdag}
q &=&-\frac{d}{d \ze} + \frac{f}{\ze} + V(\ze),\\
q^\dagger &=& \frac{d}{d\ze}  + \frac{f}{\ze} + V(\ze), \enqa
where $\ze$ is the LF invariant transverse variable and  $f$ is a
dimensionless constant.  One can add to to the Hamiltonian \req{H}
a constant term proportional to the unit matrix $\mu^2 I$ 
\beq
\label{Hmu} H_\mu =  \{Q,Q^\dagger\} + \mu^2 I,
\enq 
where the
constant $\mu$ has the dimension of a mass; thus we obtain the
general supersymmetric light-front  Hamiltonian derived in Ref. \cite{Dosch:2015bca}
 \beq \label{hmmu}
 H_\mu
  =     \left(\begin{array}{cc} - \frac{d^2}{d \ze^2}+\frac{4 L_M^2-1}{4\ze^2}+U_M(\ze)& \hspace{-1cm} 0 \\
0 & \hspace{-1cm} - \frac{d^2}{d \ze ^2} +\frac{4 L_B^2-1}{4\ze
^2}+U_B(\ze )
\end{array}\right) + \mu^2 \,\mathbf{I},
\enq where $L_B + \half = L_M - \half = f$ and $U_M$ and $U_B$
are, respectively, the meson and baryon LF confinement potentials:
\beqa
U_M(\ze) &=&    V^2(\ze) - V'(\ze) + \frac{2L_M - 1}{\ze} V(\ze),  \label{HM} \\
U_B(\ze) &=&    V^2(\ze) + V'(\ze) + \frac{2L_B+1}{\ze} V(\ze) .
\label{HB} \enqa

The superpotential $V$ is only constrained by the requirement that
it is regular at the origin.  For the special case $V= 0$, the
Hamiltonian is also invariant under conformal transformations, and
one can extend the supersymmetric algebra to a  superconformal
algebra~\cite{Akulov:1984uh,Fubini:1984hf}.  In fact, the use of
this procedure in supersymmetric quantum mechanics determines a
unique form for the superconformal potential in \req{qdag}: It is
given by $V= \sqrt{\la} \,\ze$~\cite{deTeramond:2014asa,
Dosch:2015nwa}.
Thus, in the conformal limit $\mu^2 \to 0$,  and we have \beqa
 U_M(\ze) & \to & \la_M^2 \ze^2 + 2 \, \la_M (L_M - 1),   \label{UM} \\
 U_B(\ze) &  \to &  \la_B^2 \ze^2 +  2 \, \la_B (L_B + 1),   \label{UB}
\enqa with $\la_M = \la_B = \la$. The Hamiltonian \req{hmmu} acts
on the spinor \beq \vert \phi \rangle =
\left( \begin{array}{c} \phi_M\\
\phi_B \end{array} \right) , \enq where the upper component
$\phi_M$ corresponds to a meson wave function with angular
momentum $L_M$ and a lower component $\phi_B$, which corresponds
to  the leading-twist positive chirality component of a baryon
$\psi^+$~\cite{deTeramond:2014asa, Brodsky:2014yha} with angular
momentum $L_B=L_M -1$.  The supersymmetric framework described
here also incorporates a doublet consisting of the non-leading
twist minus-chirality component $\psi^-$ of a baryon which has
angular momentum $L_B + 1$ and a its partner tetraquark with
angular momentum $L_T = L_B$~\cite{Brodsky:2016yod}. The
tetraquark sector is discussed in more detail in Ref.
\cite{Brodsky:2016yod}.

\section{Extension to the Heavy-Light Hadron Sector \label{HLS}}

In LF holographic QCD  the confinement potential for  mesons
$U_M$~\req{HM} follows from  the dilaton term $e^{\vp(x)}$
in the AdS$_5$ action following Ref.~\cite{Karch:2006pv}.  It is
given by~\cite{deTeramond:2010ge} \beq \label{dilU}
 U_{\rm dil}(\zeta)=  \frac{1}{4}(\vp'(\ze))^2  + \frac{1}{2} \vp''(\ze)
+ \frac{2L_M -3}{2 \ze} \vp'(\ze) \label{Udil}, \enq for $J_M=
L_M$. In the conformal limit a quadratic dilaton profile, $\varphi
= \lambda \zeta^2$ leads to the potential \req{UM}.

The dilaton  $\varphi$ is not constrained by the superconformal
algebraic structure in the presence of heavy quark masses,
and thus its form and the form of the superpotential  $V$ are unknown {\it a priori}. 
Additional constraints do appear, however,  by  the holographic
embedding  which can be derived by equating the potential
\req{dilU}, given in terms of the dilaton profile $\vp$, with the
meson potential \req{HM} written in terms of the superpotential
$V$.  We have: 
\beq \frac{1}{4}(\vp')^2 + \frac{1}{2} \vp''  +
\frac{2 L-1 }{2 \ze} \vp'  = V^2  - V' +\frac{2 L +1}{\ze}V, 
\enq
where $L = L_M -1$.

We shall make the ansatz: \beqa \label{a1} \vp'(\ze) &=& 2 \la \ze
\,\al(\ze), \\ \label{a2} V(\ze) &=& \la\ze \,\be(\ze) . \enqa
 Then we obtain:
 \beqa
U_{\rm dil} &=&\la^2 \ze^2 \al^2 + 2 L  \la \al + \la \ze \al' , \\
U_{\rm susy}&=&\la^2 \ze^2 \be^2 + 2 L  \la \be - \la \ze \be'  ,
\enqa and therefore \beq \label{constraint} \la^2 \ze^2
(\al^2-\be^2) + 2 L  \la (\al-\be) + \la \ze (\al' + \be')=0 .
\enq Introducing the linear combination \beqa
\si(\ze) &=& \al(\ze) + \be(\ze), \nn  \\
\de(\ze) &=& \al(\ze) - \be(\ze), \enqa it follows that \beq
 \la^2 \ze^2 \si(\ze) \de(\ze) + 2 L \la \, \de(\ze) + \la \ze \,  \si'(\ze)=0.
\enq This yields \beq \de(\ze) = - \frac{\la \ze\,
\si'(\ze)}{\la^2 \ze^2 \, \si(\ze) + 2 L \la} , \enq and
therefore: \beqa \label{a3}
\al(\ze) &=& \frac{1}{2} \left( \si(\ze)-  \frac{\la \ze\,  \si'(\ze)}{\la^2 \ze^2 \, \si(\ze) + 2 L \la}\right), \\
\label{b3} \be(\ze) &=& \frac{1}{2} \left( \si(\ze)+  \frac{\la
\ze\, \si'(\ze)}{\la^2 \ze^2 \, \si(\ze) + 2 L \la}\right) .\enqa

Using \req{a1}  and \req{a3}   we obtain  upon integration the
dilaton profile for a meson with angular momentum $L_M=L+1$ \beq
\label{dil}
 \vp(\ze) = \int d\ze\, \left(\la \ze\, \si(\ze) - \frac{ \la^2 \ze^2\,  \si'(\ze)}{\la^2 \ze^2 \, \si(\ze) + 2 (L_M-1) \la}\right).
\enq On the other hand,  from \req{a2} and \req{b3} it follows
that  this profile  for arbitrary $\si(\ze)$ is compatible with
the SUSY potential
 \beq \label{pot}
V(\ze) = \frac{1}{2} \left(\la \ze \,\si(\ze) + \frac{ \la^2 \ze^2
\, \si'(\ze)}{\la^2 \ze^2 \, \si(\ze) + 2(L_M-1) \la}\right). \enq
The baryon equations give no  further constraints.

In LFHQCD the AdS geometry fixes the nontrivial aspects of the
kinematics, whereas additional deformations of AdS space encodes
the dynamical features of the theory~\cite{deTeramond:2013it}. In
particular, the dilaton, which describes the dynamics of
confinement for mesons in holographic QCD, must be free of
kinematical quantities and thus must be independent of the angular
momentum $L_M$. This is only possible if the derivative
$\si'(\zeta) =0$ in \req{dil} and \req{pot}, thus $\si(\ze)= A$
with $A$  an arbitrary constant.  From  \req{dil} and \req{pot}
it follows  that
 \beq
  \label{V}\vp(\ze) = \half  \la A \,\ze^2 + B,  \quad \quad
  V(\ze) = \half \la A \, \ze. \label{phiV}
 \enq

This result implies that the LF potential  in the heavy-light sector, even for strongly broken
conformal invariance, has the same quadratic form as the one
dictated by the conformal algebra. The constant $A$, however, is
arbitrary, so the strength of the potential is not determined.
Notice that the interaction potential \req{Udil} is unchanged by
adding a constant to the dilaton profile, thus we can set $B =0$
in \req{phiV} without modifying the equations of motion.

The LF eigenvalue equation  $H \vert \phi\rangle =  M^2 \vert \phi
\rangle$ from the supersymmetric Hamiltonian \req{hmmu} leads to
the hadronic spectrum \beq \label{mass-formulae}
\begin{tabular}{ll}
\mbox{Mesons:}   ~~~ &  $M^2 = 4 \la_Q \, (n+L) +  \mu^2$, \\
\label{M2mu} \mbox{Baryons:}  ~~~ &  $M^2 = 4 \la_Q \, (n+L+1) +
\mu^2$,
\end{tabular}
\enq where, as we will see below, the slope constant $\la_Q =
\half  \la \, A$ can depend on the mass of the heavy quark.  The
constant   term   $\mu$  contains the effects of spin coupling and
quark masses. This term has been derived for light hadrons in
Ref.~\cite{Brodsky:2016yod}, yielding very satisfactory results,
as well as giving clear evidence for the universality of the
confinement scale $\la$ for light quarks.  More generally, we can
allow for a small breaking of the supersymmetry due to the
different light quark masses in the meson or nucleon, $\mu^ 2_M
\simeq \mu^2_B \simeq \mu^2$. We shall discuss a possible
extension for heavy quarks in the Appendix~\ref{app}, but we will initially 
treat their masses  as unconstrained constants in a fit to  all the
heavy-light trajectories.

\section{Comparisons with Data \label{exp}}

\begin{figure}[ht]
\begin{center}
\includegraphics[width=8cm]{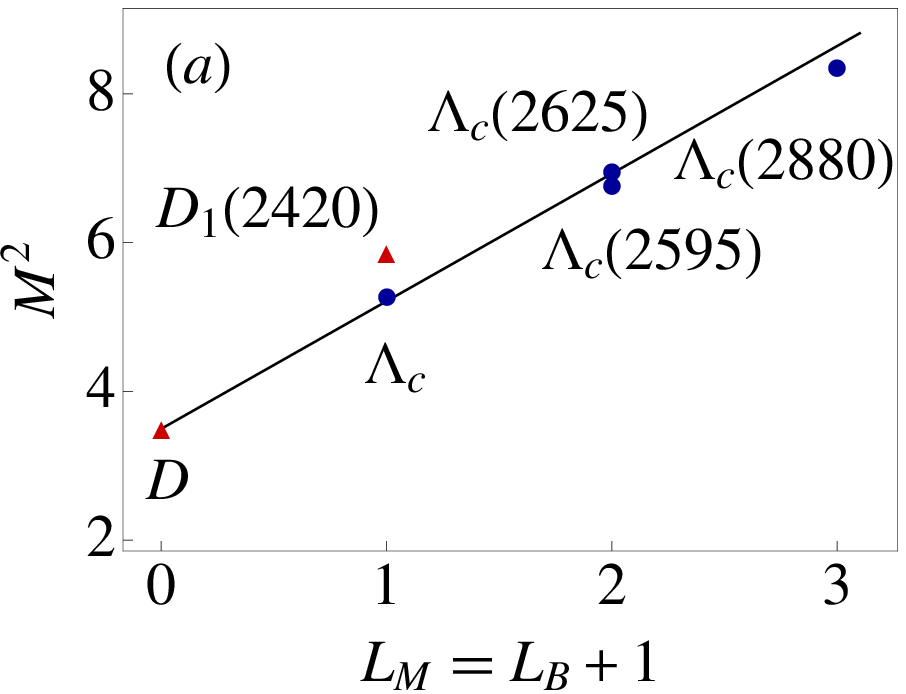}
\includegraphics[width=8cm]{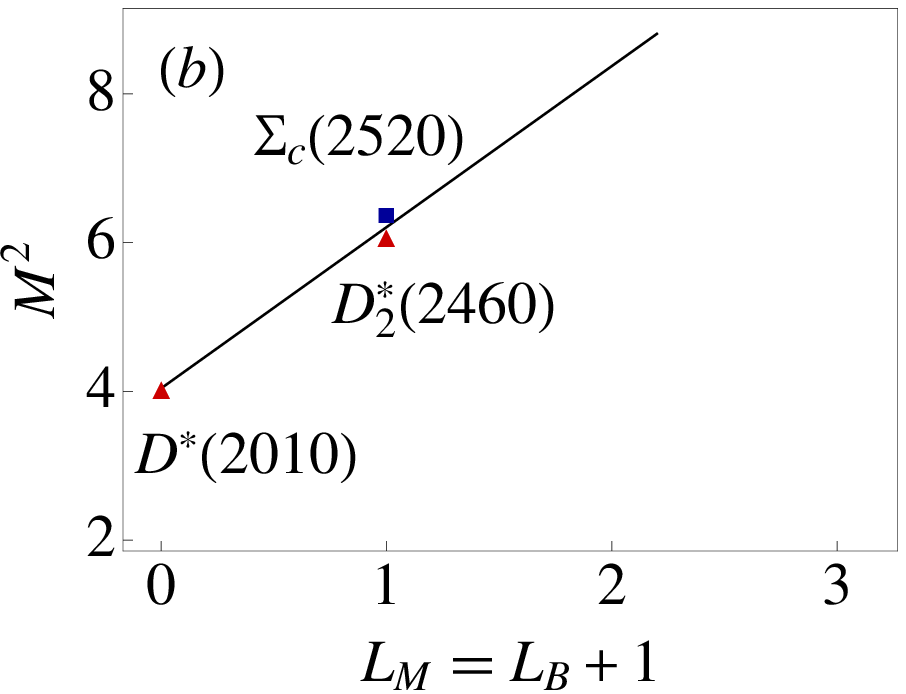}
\includegraphics[width=8cm]{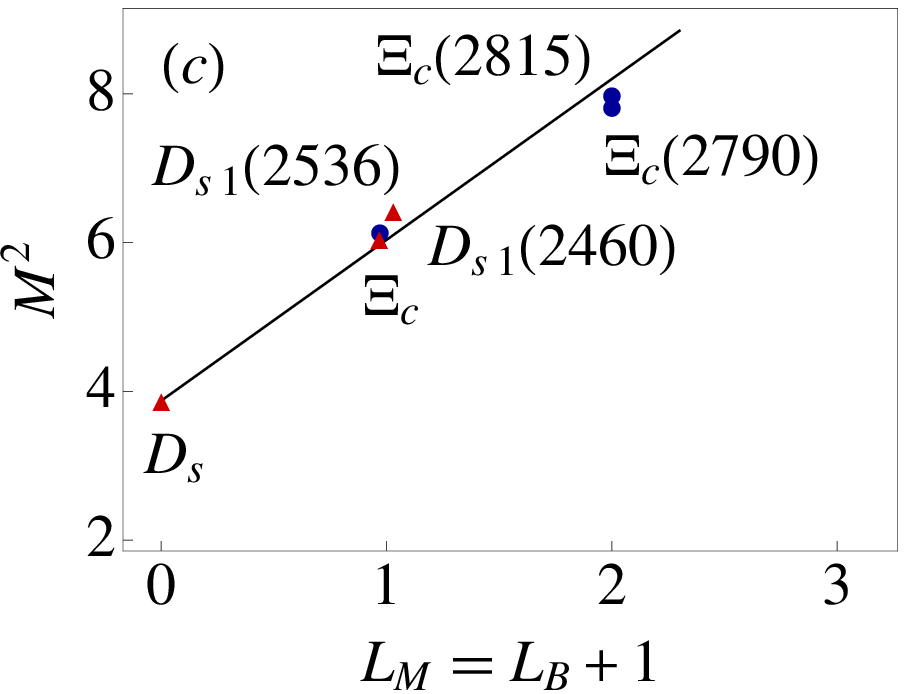}
\includegraphics[width=8cm]{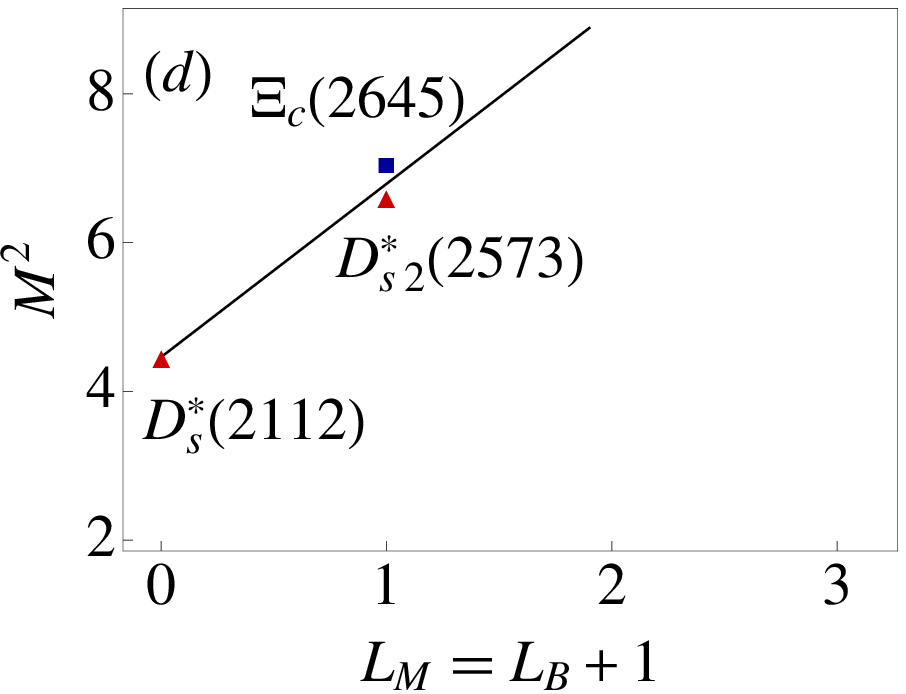}
\end{center}
\caption{\label{figcharm} Heavy-light mesons and baryons
with one charm quark: $D= q \bar c$, $D_s = s \bar c$, $\Lambda_c
= u d c$, $\Si_c = q q c$, $\Xi_c = u s c$. In (a) and (c) $s=0$
and in (b) and (d) $s =1$, where $s$ is the total quark spin in
the mesons or the spin of the quark cluster in the baryons. The
data is from Ref. \cite{Olive:2016xmw}.}
\end{figure}

\begin{figure}[ht]
\begin{center}
\includegraphics[width=8cm]{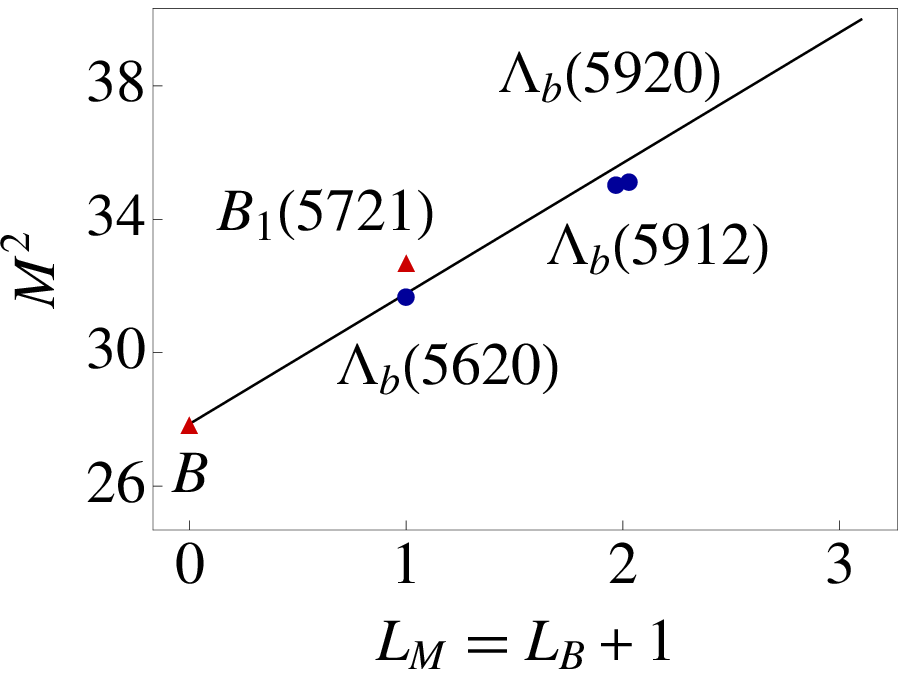}
\includegraphics[width=8cm]{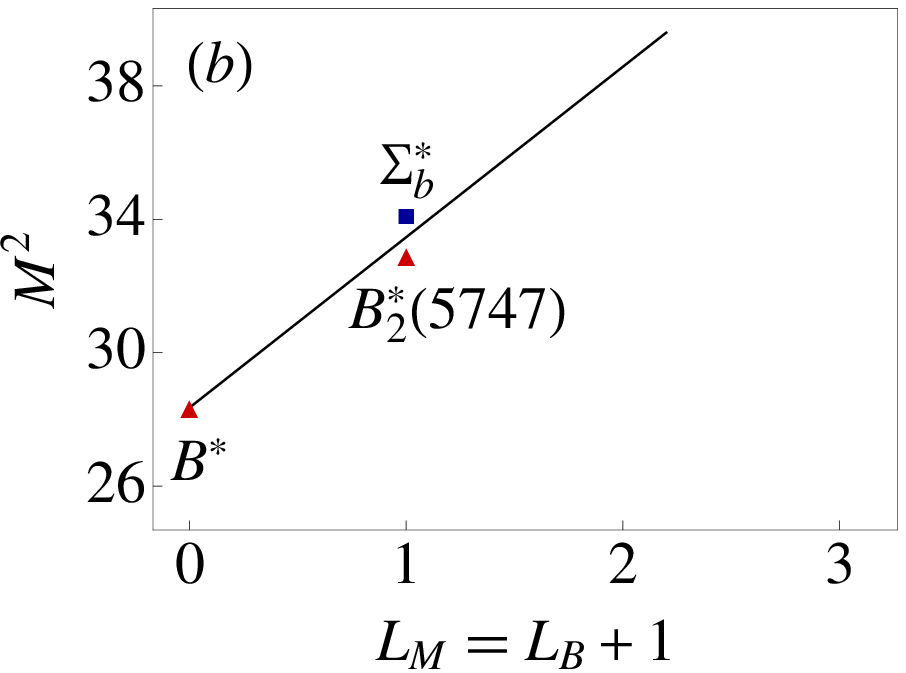}
\includegraphics[width=8cm]{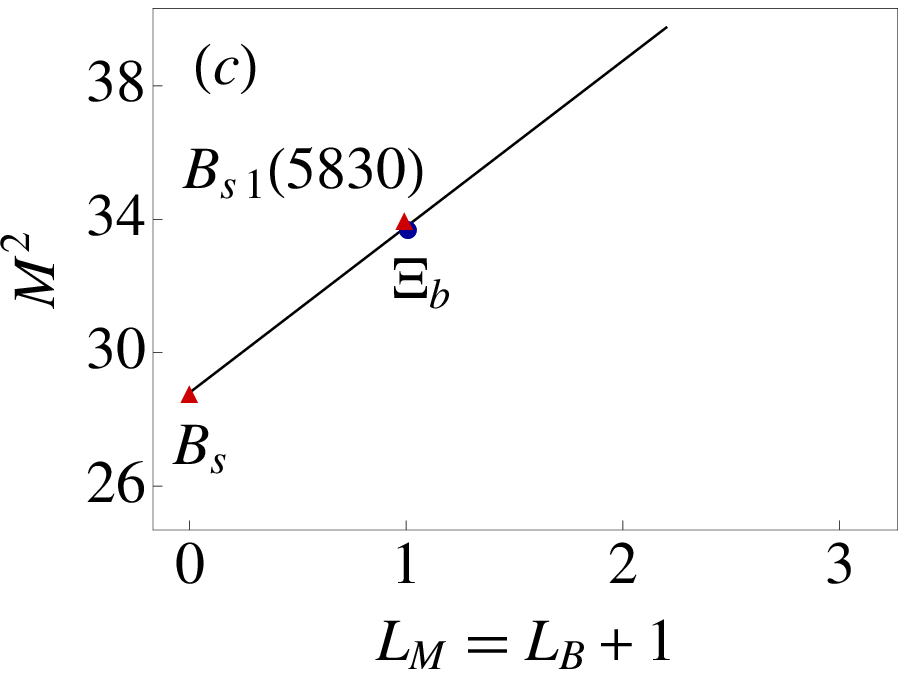}
\includegraphics[width=8cm]{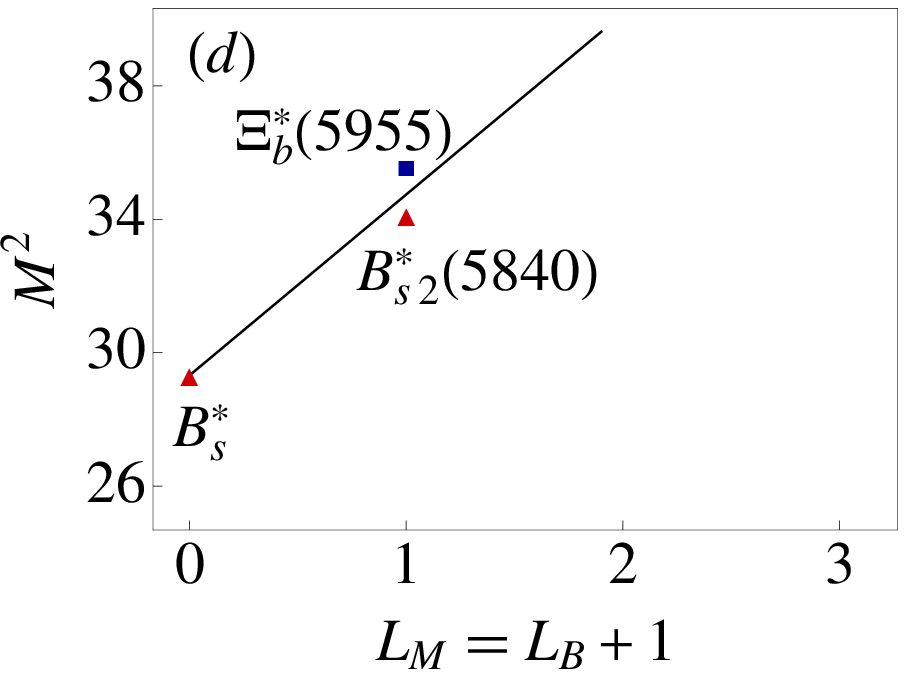}
\end{center}
\caption{\label{figbottom} Heavy-light  mesons and baryons
with one bottom quark:   $B= q \bar b$, $B_s = s \bar b$,
$\Lambda_c = u d b$, $\Si_b = q q b$, $\Xi_c = u s b$. In (a) and
(c) $s=0$ and in (b) and (d) $s =1$, where $s$ is the total quark
spin in the mesons or the spin of the diquark cluster in the
baryons. The data is from Ref. \cite{Olive:2016xmw}.}
\end{figure}

In Figs.~\ref{figcharm} and \ref{figbottom} we display
confirmed data for the heavy-light mesons and baryons containing
one charm or one bottom quark together with the trajectory fit
from  \req{mass-formulae}.  The internal spin $s$ in these figures
refers to the total quark spin in the mesons or the spin of the
diquark cluster in the baryons~\cite{Brodsky:2016yod}. The results
presented in Figs. \ref{figcharm} and \ref{figbottom} constitute a
test of  the linearity of the trajectories predicted by the SUSY
holographic embedding, and it allows us to determine the
dependence of the slope $\la_Q$ on the heavy quark mass scale. The
trajectory intercepts are fixed by the lowest state in each
trajectory, but are determined later by the model in
the  Appendix~\ref{app}. Unfortunately the data for  heavy-light
hadrons are sparse, compared with those for light hadrons. Only
the $D/\La_c$ trajectory, Fig.~\ref{figcharm} (a)  provides an
independent test for the predicted harmonic potential. Thus,
future data on heavy-light hadrons will be essential  to test
the assumptions  stated in Sec.~\ref{intro} for the light front
holographic model described here.

\begin{figure}
\begin{center}
\includegraphics*[width=12cm]{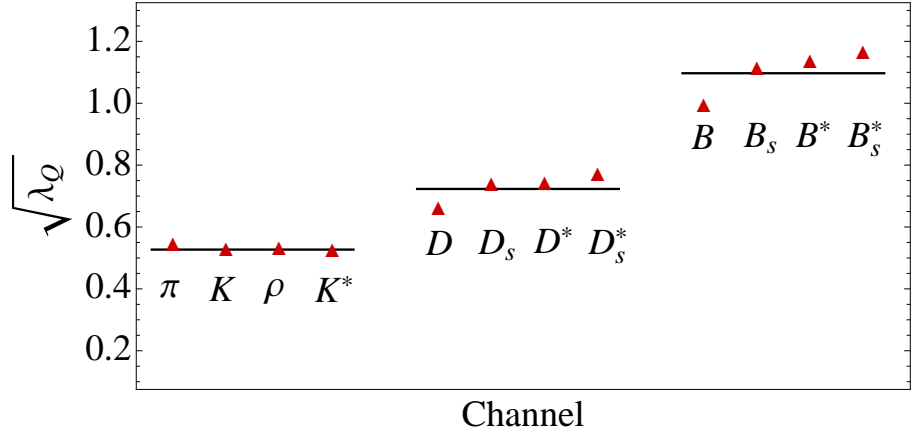}
\end{center}
\caption{\label{lambda-channel} The fitted value of $\sqrt{\la_Q}$
for different meson-baryon trajectories, indicated by the lowest
meson state on that trajectory.}
\end{figure}

In Fig.~\ref{lambda-channel} the fitted values for $\sqrt{\la_Q}$
are presented for the different trajectories.  In the abscissa we
indicate the lowest mass meson for that meson-baryon trajectory.
The triangles indicate the fitted values, and the horizontal lines
show the mean over all channels of hadrons containing the same
heavy-light meson. For comparison, we also give the
corresponding values for a fit to the much more abundant data for
light hadrons~\cite{Brodsky:2016yod}. It is obvious that the
dispersion of the data is significantly smaller  for the case
where  the model is approximately constrained by conformal
symmetry, as compared to the case where it is strongly broken by
heavy quark masses, and only supersymmetry remains as a constraint.

All of the results for the charmed hadrons are collected in Table
\ref{tabcharm};  the predictions for bottom hadrons  are
summarized in Table \ref{tabbottom}.   The slopes for charm
hadrons are definitely larger than those for the light hadrons,
but they agree within $\pm 10 \% $ for  all charm hadrons. The
agreement of the data with the theoretical  predictions from
\req{mass-formulae} is of the same order as  for light hadrons.
The average deviation is 55 MeV, but the data are rather sparse.
The model, however, makes predictions for higher orbital (and
radial) excitations with an accuracy of approximately $\pm 100$
MeV. The values for the mean of the modulus of deviation between
theoretical and experimental values is 55 MeV,  the standard error
is 72 MeV; this deviation is  comparable to that obtained for
light hadrons  \cite{Dosch:2015bca,Brodsky:2016yod}.  We have
added in Table \ref{tabcharm} the predicted missing superpartners
and all mesons with angular momentum $L_M \leq 2$ and baryons with
$L_B \leq 1$.

We have omitted the $\Si_c$ and the $\Si_b$ baryons from the figures
and the tables, since it is not clear whether  they should be
included in the same trajectories with the pseudoscalar or the vector
meson, as will be discussed in more detail at the end of  the
Appendix~\ref{app}.

\begin{table}
\caption{\label{tabcharm}  Charmed Hadrons. The quark spin $s$ is
the total quark spin of the meson or the diquark cluster, $\la_Q$
is the fitted value for the  trajectory and $\Delta M$ is the
difference between the observed and the theoretical value
according to \req{mass-formulae}. The lowest lowest lying meson
mass determines de value of  $\mu^2$ in \req{M2mu} for each
trajectory. We have added predictions, if only one superpartner
has been observed and for $L_M\leq 2,\;L_B\leq 1$.}
\begin{center}
{\renewcommand{\arraystretch}{0.7}
\begin{tabular}{cccccccc}
status&particle&$I(J^P)$&quark&spin&$n,L$&$\sqrt{\la_Q}$&$\Delta M$\\
&&& content&&&[GeV]&[MeV]\\
\hline
obs& $D(1869)$&$\half(0^-)$&$c \bar q$&0&$0,0$&0.655& 0\\
obs&$D_1(2400)$&$\half(1^+)$&$c \bar q$&0&$0,1$&0.655&139 \\
obs&$\La_c(2286)$&$0(\half^+)$&$c q q$&$0$&$0,0$&0.655&4\\
obs&$\La_c(2595)$&$0(\half^-)$&$c q q$&$0$&$0,1$&0.655&-36\\
obs&$\La_c(2625)$&$0(\thalf^-)$&$c q q$&$0$&$0,1$&0.655&-6\\
obs&$\La_c(2880)$&$0(\fhalf^+)$&$c q q$&$0$&$0,2$&0.655&-59\\
pred& $D_2({\it 2630})$&$\half(2^-)$&$c \bar q$&0&$0,2$&0.655&?\\
pred& $D_2({\it2940})$&$\half(3^+)$&$c \bar q$&0&$0,3$&0.655& ?\\
\hline
obs& $D^*(2007)$&$\half(1^-)$&$c \bar q$&1&$0,0$&0.736& 0\\
obs& $D_2^*(2460)$&$\half(2^+)$&$c \bar q$&1&$0,1$&0.736& -29\\
obs&$\Si_c(2520)$&$1(\thalf^+)$&$c q q$&$1$&$0,0$&0.736&28\\
pred& $D_3^*({\it 2890})$&$\half(3^-)$&$c \bar q$&1&$0,2$&0.736& ?\\
pred&$\Si_c({\it 2890})$&$1(\fhalf^-)$&$c q q$&$1$&$0,1$&0.736&?\\
pred&$\Si_c({\it 2890})$&$1(\thalf^-)$&$c q q$&$1$&$0,1$&0.736&?\\
pred&$\Si_c({\it 2890})$&$1(\half^-)$&$c q q$&$1$&$0,1$&0.736&?\\
\hline
obs& $D_s(1958)$&$0(0^-)$&$c \bar s$&0&$0,0$&0.735& 0\\
obs&$D_{s1}(2460)$&$0(1^+)$&$c \bar s$&0&$0,1$&0.735&23\\
obs&$D_{s1}(2536)$&$0(1^+)$&$c \bar s$&0&$0,1$&0.735&73\\
obs&$\Xi_c(2467)$&$\half(\half^+)$&$c s q$&$0$&$0,0$&0.735&31\\
obs&$\Xi_c(2575)$&$\half(\half^+)$&$c s q$&$0$&$0,0$&0.735&113\\
obs&$\Xi_c(2790)$&$\half(\half^-)$&$c s q$&$0$&$0,1$&0.735&-67\\
obs&$\Xi_c(2815)$&$\half(\thalf^-)$&$c s q$&$0$&$0,1$&0.735&-41\\
pred& $D_{s2}({\it 2856})$&$0(2^-)$&$c \bar s$&0&$0,2$&0.735& ?\\
\hline
obs&$D^*_s(2112)$&$0(1^-) ?$&$c \bar s$&1&$0,0$&0.766& 0\\
obs&$D^*_{s2}(2573)$&$0(2^+) ?$&$c \bar s$&1&$0,1$&0.766& -29\\
obs&$\Xi_c(2646)$&$\half(\thalf^+)$&$c s q$&$1$&$0,0$& 0.766&28\\
obs&$D^*_{s3}({\it 3030})$&$0(3^-) ?$&$c \bar s$&1&$0,2$&0.766& 0\\
pred&$\Xi_c({\it 3030})$&$\half(\fhalf^-)$&$c s q$&$1$&$0,1$& 0.766&?\\
pred&$\Xi_c({\it 3030})$&$\half(\thalf^-)$&$c s q$&$1$&$0,1$& 0.766&?\\
pred&$\Xi_c({\it 3030})$&$\half(\half^-)$&$c s q$&$1$&$0,1$& 0.766&?\\
\end{tabular}}
\end{center}
\end{table}

\begin{table}
\caption{\label{tabbottom}  Bottom Hadrons. The notation is the
same as for Table. \ref{tabcharm}.}

\begin{center}
{\renewcommand{\arraystretch}{0.7}
\begin{tabular}{cccccccc}
status&particle&$I(J^P)$&quark&spin&$n,L$&$\sqrt{\la_Q}$&$\Delta M$\\
&&& content&&&[GeV]&[MeV]\\
\hline
obs& $B(5279)$&$\half(0^-)$&$b \bar q$&0&$0,0$&0.963& 0\\
obs&$B_1(5721)$&$\half(1^+)$&$b \bar q$&0&$0,1$&0.963&101 \\
obs&$\La_b(5620)$&$0(\half^+)$&$b q q$&$0$&$0,0$&0.963&1\\
obs&$\La_b(5912)$&$0(\half^-)$&$b q q$&$0$&$0,1$&0.963&-28\\
obs&$\La_c(5920)$&$0(\thalf^-)$&$b q q$&$0$&$0,1$&0.963&-20\\
pred& $B_2({\it 5940})$&$\half(2^-)$&$c \bar q$&0&$0,2$&0.963&?\\
\hline
obs& $B^*(5325)$&$\half(1^-)$&$b \bar q$&1&$0,0$&1.13& 0\\
obs& $B_2^*(5747)$&$\half(2^+)$&$b \bar q$&1&$0,1$&1.13& -45\\
obs&$\Si^*_b(5833)$&$1(\thalf^+)$&$b q q$&$1$&$0,0$&1.13&44\\
pred& $B_3^*({\it 6216})$&$\half(3^-)$&$c \bar q$&1&$0,2$&1.13& ?\\
pred&$\Si_b({\it 6216})$&$1(\fhalf^-)$&$c q q$&$1$&$0,1$&1.13&?\\
pred&$\Si_b({\it 6216})$&$1(\thalf^-)$&$c q q$&$1$&$0,1$&1.13&?\\
pred&$\Si_b({\it 6216})$&$1(\half^-)$&$c q q$&$1$&$0,1$&1.13&?\\
\hline
obs& $B_s(5367)$&$0(0^-)$&$b \bar s$&0&$0,0$&1.11& 0\\
obs&$B_{s1}(5830)$&$0(1^+)$&$b \bar s$&0&$0,1$&1.11&16\\
obs&$\Xi_b(5795)$&$\half(\half^+)$&$b s q$&$0$&$0,0$&1.11&-16\\
pred& $B_{s2}({\it 6224})$&$0(2^-)$&$b \bar s$&0&$0,2$&1.11& ?\\
pred &$\Xi_b(\it 6224)$&$\half(\half^-)$&$b s q$&$0$&$0,1$&1.11&?\\
pred &$\Xi_b(\it 6224)$&$\half(\thalf^-)$&$b s q$&$0$&$0,1$&1.11&?\\
\hline
obs&$B^*_s(5415)$&$0(1^-) ?$&$b \bar s$&1&$0,0$&1.16& 0\\
obs&$B^*_{s2}(5840)$&$0(2^+) ?$&$b \bar s$&1&$0,1$&1.16& -55\\
obs&$\Xi_b(5945)$&$\half(\thalf^+)$&$b s q$&$1$&$0,0$& 1.16&55\\
pred&$B^*_{s3}(6337)$&$0(3^-) ?$&$b \bar s$&1&$0,2$&1.16& ?\\
pred&$\Xi_b({\it 6337})$&$\half(\fhalf^-)$&$b s q$&$1$&$0,1$&1.16& ?\\
pred&$\Xi_b({\it 6337})$&$\half(\thalf^-)$&$b s q$&$1$&$0,1$&1.16& ?\\
pred&$\Xi_b({\it 6337})$&$\half(\half^-)$&$b s q$&$1$&$0,1$&1.16& ?\\
\end{tabular}}
\end{center}

\end{table}

\section{The Scale Dependence of $\la_Q$ from  Heavy Quark Effective Theory (HQET) \label{sechqet}}

It has been known for a long time \cite{Shuryak:1981fza}, and has been formally proved in HQET \cite{Isgur:1991wq}, that in the case of masses of heavy mesons $M_M$, the product $\sqrt{M_M\,}\,f_M$ approaches, up to logarithmic terms,  a finite value 
\beq
\label{hqet} \sqrt{M_M\,} \, f_M \to C, 
\enq
a relation which can also be derived using the light-front holographic approach~\cite{Branz:2010ub}. In the present holographic framework this means that the confinement scale $\la_Q$ has to increase with increasing quark mass. Indeed, using the results of the Appendix~\ref{app}, we can write  the decay constant $f_M$ \req{fmapp} expressed through the wave function \req{wf} 
\beq \label{fmmod} 
f_M= \frac{1}{\sqrt{\int_0^1
dx\,  e^{ - m_Q^2 /\la(1-x)}}}{\frac{\sqrt{2 N_C\la}}{\pi} }\int_0^1
dx\,  e^{- m_Q^2/2 \la (1-x)} \sqrt{x(1-x)},
\enq
where, for simplicity, we consider the case where $m_1=0$; the heavy quark mass is  $m_2=m_Q$.

\begin{figure}[ht]
\includegraphics[width=12cm]{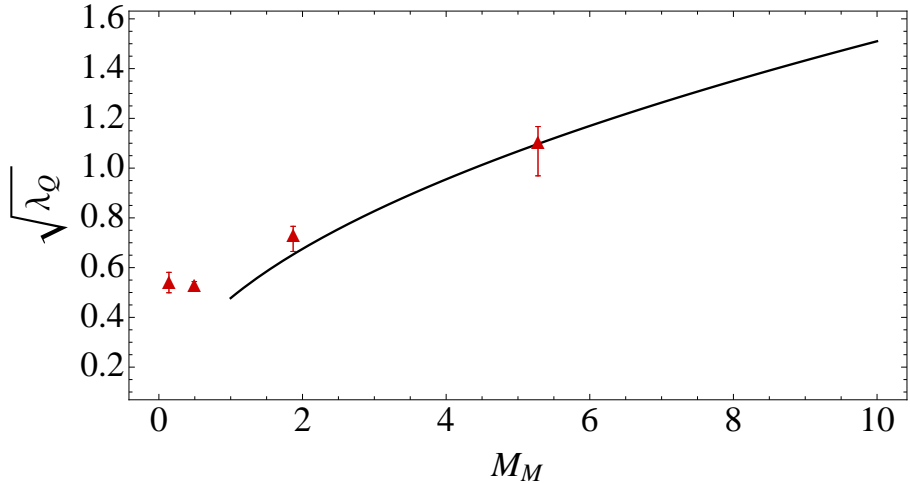}
\caption{\label{lambdamass} The fitted value of $\la_Q$ vs. the
meson mass $M_M$. The solid line is the square root dependence
\req{hq} predicted by HQET.}
\end{figure}

We introduce $\nu^2 \equiv m_Q^2/\la$ and use the saddle-point method to evaluate the integral of the numerator for large values of $\nu^2$. One expands the numerator around the value $x_0= \frac{1}{\nu^2} +O\left(\frac{1}{\nu^4}\right) $, where the integrand is maximal and  obtains: 
\beq e^{-\half
\nu^2/(1-x)}\sqrt{x(1 -x)} = e^{-\nu^2/2 - \log \nu -1/2
+O\left(\frac{1}{\nu}\right) } \, e^{\frac{1}{4}
(x-x_0)^2\,\left(m^4 +O\left( \nu^2 \right)\right)} .
\enq 
This Gaussian  integral  yields: 
\beq \int_0^1 d x \, e^{-\half
\nu^2/(1-x)}\sqrt{x(1-x)}=\frac{ e^{-\nu^2/2}}{\sqrt{e\,
\nu^2} }\frac{\pi}{\nu^2}\left(1+{\rm erf}\left( \half \right)
\right).
\enq 
The integral in the denominator of \req{fmmod} can be performed analytically 
\beq \int_0^1 dx\,  e^{
-\nu^2/(1-x)}=\int_1^\infty \frac{dy}{y^2} \,e^{-\nu^2\,y}
=e^{-\nu^2} - \nu^2\,\Gamma \left(0,\frac{1}{\nu^2}\right)
=e^{-\nu^2}\left( \frac{1}{\nu^2} +
O\left(\frac{1}{\nu^4}\right)\right).
\enq
 Thus in the  large $m_Q$ limit:
\beq 
\label{fmhqet} f_M = \sqrt{\frac{6}{ e}} \left(1+{\rm
erf}\left(\half \right) \right) \frac{\la^{3/2}} {m_Q^2} .
\enq

In the limit of heavy quarks the meson mass equals  the quark mass. From the HQET relation \req{hqet} it  follows that 
\beq
\label{hq} 
\la_Q = {\rm const}  ~ m_Q,
\enq 
where the constant in \req{hq} has the dimension of mass.
This corroborates our statement that the increase of $\la_Q$ with increasing quark mass is dynamically necessary. In Fig. \ref{lambdamass} we show the value of $\lambda_Q$ for the $\pi,\,K,\,D,$ and $B$ mesons  as function  of the meson mass $M_M$. From the difference of the values of $\sqrt{M_M}\,f_M$ for the $D$ and $B$  mesons (see Appendix \ref{app}, Table \ref{leptonic}) we must conclude that, in this region, we are still far away from the heavy quark regime. It is nevertheless remarkable  that the simple functional dependence \req{hq} derived in the heavy quark limit predicts for the $c$ quark a value $\sqrt{\la_c}=  0.653$  GeV --~after fixing the proportionality constant in \req{hq} at the B meson mass, which is indeed at the lower edge of the values obtained from the fit to the trajectories (0.655 to 0.766  GeV).  It makes no sense  to apply HQET below the mass of the $M_D$. Indeed, there is no sign of an increase of $\sqrt{\la}$  between the $\pi$ and $K$ mass.

\section{Summary and Conclusions \label{sc}}

In this article we have  extended  light-front holographic QCD to heavy-light hadrons by using the  embedding of supersymmetric quantum mechanics  in a modified higher dimensional space asymptotic to AdS. Remarkably,  this embedding not only yields supersymmetric relations between mesons and baryons, but it also  determines the superconformal potential and thus the effective potential in light-front holographic QCD. If one introduces for mesons the breaking of the maximal symmetry  of  AdS$_5$ by a dilaton term, as it is usually done, one finds that only a quadratic dilaton profile is compatible with the supersymmetric potential; thus, a harmonic LF potential again emerges, as is the case for light quark hadrons. This implies linear trajectories not only for light hadrons, but also for the heavy-light mesons and baryons. Although the experimental data are sparse, the existing data are not in contradiction with this linearity; however, future data on heavy-light hadrons will be critical to test the dynamical assumptions described here.

In our approach, the heavy  quark  influences the transverse degrees of freedom only indirectly by modifying  the strength of the harmonic potential; this modification cannot be determined from supersymmetry. However, the dependence of the confinement scale on the heavy quark mass can be calculated in HQET, and it is in agreement with the observed increase. Indeed, HQET is compatible with the light front holographic approach to hadron physics~\cite{Branz:2010ub}.

\acknowledgements

S.J.B. is supported by the Department of Energy, contract DE--AC02--76SF00515. SLAC-PUB--16882.

\appendix

\section{Wave functions and distribution amplitudes \label{app}}

As mentioned above, the additional term $\mu^2$  in
Eq.~\req{mass-formulae}  for light hadrons was given in
\cite{Brodsky:2016yod}  in terms of the internal spin and the
quark masses of the constituents. The spin interaction term
has the simple form  $2 \la \,  s$, where $s$ is the quark spin of
the meson or the quark spin of the diquark cluster in the
baryon, respectively.  There is, however a problem with the
cluster spin assignment of the $\Si_c$ and $\Si_b$, as will be
explained at the end of this appendix.

In order to estimate the influence of the quark masses and also to
evaluate the decay constants $f_M$, which play a crucial role in
Sec.~\ref{sechqet}, we need to have a good description of
the wave functions of the
hadrons. We found for a hadron with LF angular momentum $L$ and
radial excitation number $n$~\cite{Brodsky:2014yha}:
\beq \label{wf0}  \psi^{(0)}_{n,L}=\frac{1}{N}\sqrt{x(x-1)} \, 
\zeta^L  L_n^L(|\la| \ze^2) \,e^{-|\la| \ze^2/2},
\enq
with normalization
\beq 
N=\sqrt{\frac{(n+L) !}{\, n!\, \pi}}\, |\la|^{(L+1)/2}.
\enq 
Here $L_n^L$ are the associated Laguerre Polynomials, and $\ze=\sqrt{x_1(1-x_1) }\,| b_{\perp1}|$ for
mesons and $\ze= \sqrt{\frac{x_1}{1-x_1}}\, |\,(x_2 b_{\perp
2}+x_3 b_{\perp 3})\,| $  for baryons; $b_{\perp i}$ is the
transverse distance of quark $i$  from the impact line defined by
$\sum_{i=1}^n b_{\perp i}=0$.

LFHQCD gives us no hints on the longitudinal dynamics, so we  have
constructed the wave function for hadrons with light quarks of
mass $m_i$ by the principle, that the wave function is
determined by the invariant mass of the constituents \beq
\sum_{i=1}^n \frac{k_{\perp i}^2 + m_i^2}{x_i}, \enq where
$k_{\perp i}$ is the transverse momentum of the constituent $i$.
This leads to the wave function for hadrons with small quark
masses: \beq \label{wf} \psi^{(m)}_{n,L}= \frac{1}{N_m} \,
e^{-\frac{1}{2\la} \Delta m^2} \, \psi^{(0)}_{n,L},
 \enq
with
 \beq  \label{invmass}
 \Delta m^2=\sum_{i=1}^n \frac{m_i^2}{x_i} \, \de\Big(\sum_{i=1}^nx_i-1\Big) . 
 \enq
The normalization condition  $\int_0^1  dx_1 \cdots dx_n \,
\de\Big(\sum_{i=1}^nx_i-1\Big)  \int  d^2b_\perp\,
|\psi^{(0)}_{n,L}|^2  =1$ implies
 \beq
  N_m^2 =\int_0^1  dx_1 \cdots dx_n \, \de\Big(\sum_{i=1}^nx_i-1\Big) \, e^{-\frac{1}{\la} \Delta m^2}.
\enq

It is certainly not realistic  to assume that these  wave
functions, derived  under the assumption of small quark masses, can
be  simply extrapolated to heavy-light hadrons. But on the other
hand, the embedding of the supersymmetric theory into modified AdS
demands that the quark masses  enter only indirectly through the confining
(transverse) dynamics, namely by a change of the confinement scale
$\lambda$. We therefore apply, in an exploratory way, the procedure 
developed for light quarks~\cite{Brodsky:2014yha}
 to determine also the masses  of hadrons containing a heavy quark.

According to \cite{Brodsky:2016yod} the  set of constants $\mu^2$ in
\req{mass-formulae} are given in first approximation by: 
\beq
\mu^2= 2 \la \,s +\Delta M^2[m_1,  \cdots, m_n] ,
\enq
where the first term is the spin term discussed above and
 \beq 
 \Delta M^2[m_1, \cdots, m_n] = \int 2 \pi \ze d\ze 
 \int dx_1 \cdots dx_n\, \psi(\zeta,x_1, \cdots, x_n)^2\,\sum_{i=1}^n \frac{m_i^2}{x_i} \, \de\Big(\sum_{i=1}^nx_i-1\Big) ,
\enq 
where $\psi$ is the normalized ground state wave function \req{wf} with
 $n=2$ for mesons and $n=3$ for baryons.

Since  $\la_Q$ has been determined in the fit to the trajectories
and the light quark masses are known from the fits to light
hadrons~\cite{Brodsky:2014yha}, the only free parameter in these
formul\ae \, is the effective heavy quark mass, $m_Q$. For
hadrons containing a charm quark, the best fit to the 8 ground
states of the trajectories yields $m_c = 1547$ MeV, for the bottom
quark mass one obtains correspondingly $m_b =  4922$ MeV. The quality
of the fit is worse than  that to the trajectories, the standard
deviation is 95 MeV.

A more severe test for the adequacy of the wave functions are the
leptonic decay constants. The leptonic decay constant of a
pseudoscalar meson $M$ samples the light-front wave function at
small distances and is a very sensitive test for the wave
function. Its exact computation is given in terms of the valence
light-front wave function~\cite{Lepage:1980fj, Brodsky:2007hb}
\beq \label{fmapp}
 f_M = 2 \sqrt{2 N_C} \int_0^1 dx  \, \phi(x), \enq where \beq \label{DA}
\phi(x) = \int \frac {d^2 {\mbf k}_\perp}{16 \pi^3}  \, \psi(x,
\mbf{k}_\perp), 
\enq
is the distribution amplitude (DA). Since
$\phi(x) = \psi(x, \mbf{b}_\perp = 0)/  \sqrt{4 \pi}$, we can
write $f_M$ in terms of the LFWF at zero transverse impact
distance: 
\beq \label{fm2} f_M=\sqrt{\frac{2
N_C}{\pi}}\int_0^1dx\, \psi(x, \mbf{b}_\perp = 0),
\enq 
which is identical with the result first obtained by van Royen and
Weisskopf~\cite{VanRoyen:1967nq}.

The  decay constants $f_M$ of the heavy-light mesons are
not directly observable, since the leptonic decay rates also
depend on the matrix elements of the weak decay of heavy quarks.
There are, however, many phenomenological results, notably from
QCD sum rules and lattice calculations, which give a fairly
consistent picture. We present in Table \ref{leptonic}, second row,
the results form \cite{Olive:2016xmw}, {\it Leptonic decays of
charged pseudoscalar mesons}.  For completeness we have also
included the $K$ meson.

\begin{table}
\caption{\label{leptonic} Leptonic decay constants. Second row:
the phenomenological values; third row: theoretical values
obtained from  \req{fm2} with the unmodified wave function \req{wf} and the
fitted heavy quark masses  $m_c =  1547, \;m_b=4922$ MeV; last row:
theoretical values obtained with the modified wave function with  
the scale factor $\al=\half$  in \req{modfac}. The fitted masses are $m_c = 1327, \; m_b = 4572$ MeV.}
\renewcommand{\arraystretch}{0.7}
\bec
\begin{tabular}{lccccc|cc}
decay const. [MeV] &$f_K$&$f_D$&$f_{Ds}$&$f_B$&$f_{B_s}$&$\frac{f_{D_s}}{f_{D}}$&$\frac{f_{B_s}}{f_B}$\\
\hline
phenomenology&155&212&249&187&227&1.17&1.22\\
unmodified w.f. & 152 & 127&  159&  81& 117& 1.25& 1.44\\
modified w.f. &-& 199&  216& 194&  229& 1.09& 1.18\\
\hline
\end{tabular}
\enc
\end{table}

The results for the decay constants obtained from \req{fm2} with
the wave function \req{wf} are displayed in Table~\ref{leptonic},
third row, ``unmodified w.f.''. Though qualitative features are
reproduced,  the magnitude of the decay constants is grossly
underestimated with increasing heavy quark mass. This is due to
the fact that the heavy quark carries most of the longitudinal
momentum, as it is formally expressed through the $x_i$ dependent
exponent $\De m^2$  \req{invmass} in \req{wf}.  If the heavy quark
mass  $m_2$ increases, then  $x_1$ is pushed to very small values;
this suppresses  the decay constant $f_M$. Since this suppression is
evidently too strong, an easy remedy is to multiply the heavy quark
mass in the exponential \req{invmass} of the wave function
\req{wf} by a factor $\alpha <1$; thus we modify
\beq \label{modfac}
e^{- \frac{1}{2 \la} \frac{m_Q^2}{x_Q}} \to e^{- \frac{\al^2}{2 \la} \frac{m_Q^2}{x_Q}},
\enq
in the LF wave function for the heavy quark with mass $m_Q$ and longitudinal momentum $x_Q$.

The result for $\alpha= \half$
is shown in Table~\ref{leptonic}, last row, ``modified w.f.". The improvement
from  errors between 40\% and 60\% to errors  between 3\% and 8\% is dramatic, and, most
important, there is no sign of an increasing discrepancy with
increasing quark mass. Since the quantity $\alpha$ is mass
independent, it does not affect the conclusions from HQET,  drawn
in Sec.~\ref{sechqet}, notably the relation \req{hq}; only the
value of $m_Q$ in  \req{fmhqet} has to be multiplied by
$\al=\half$. The values for the quark masses, obtained from a fit
to the data with this modified wave function are: $m_c = 1.327$ GeV
and $m_b =  4.572$ GeV. The fit is slightly worse than that with the
unmodified wave function \req{wf}, the standard deviation is 125
MeV.

\begin{figure}[]
\includegraphics*[width=8.2cm]{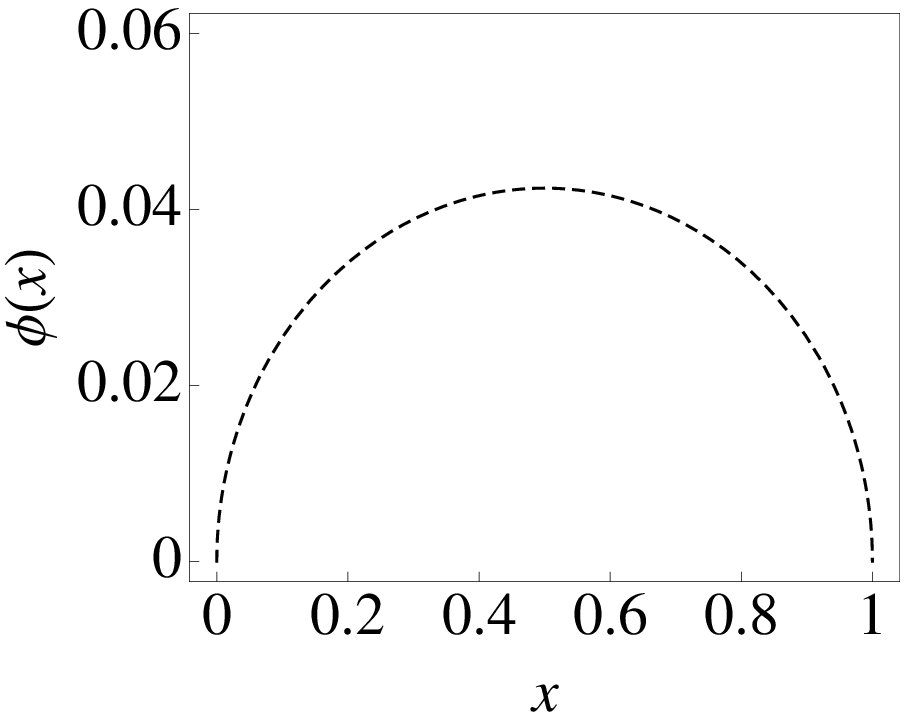}\\
\includegraphics*[width=8.2cm]{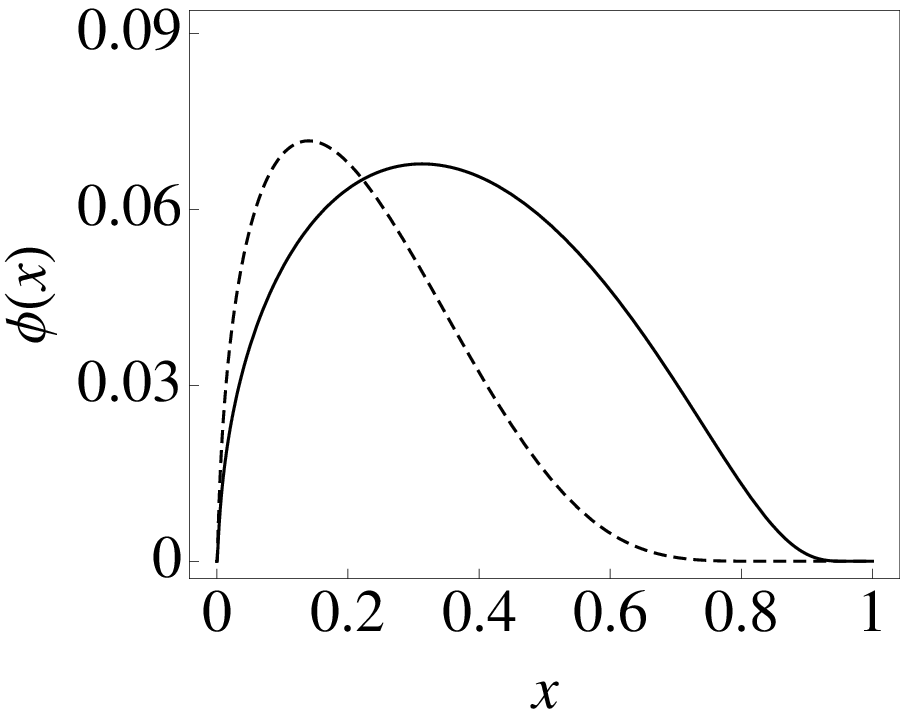}
\includegraphics*[width=8.2cm]{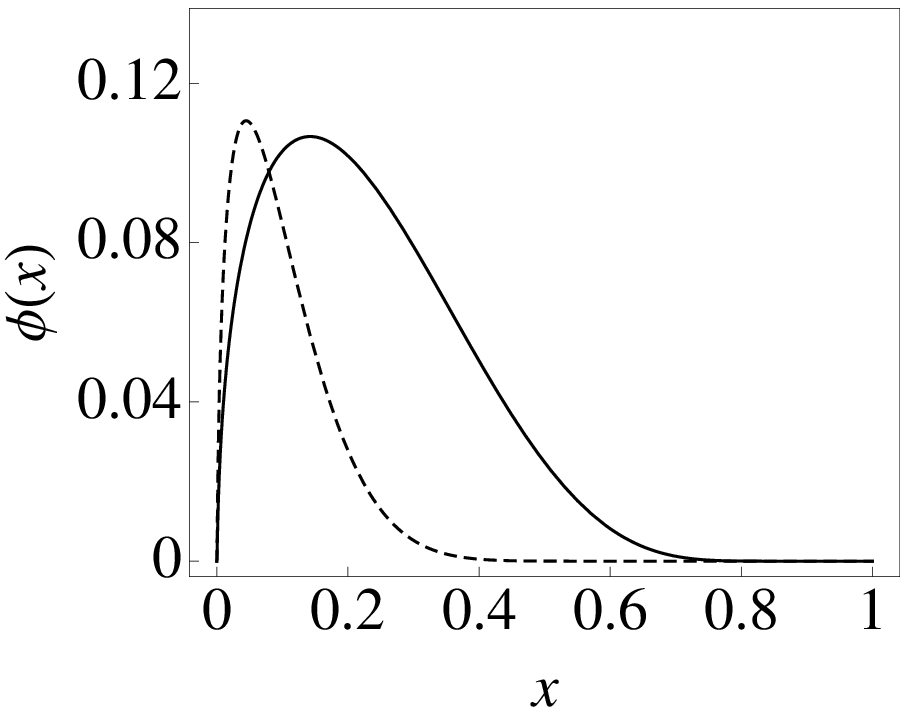}
\caption{\label{DA} Distribution amplitudes for pseudoscalar
mesons. From top to bootm: Chiral case, D meson and B meson. The
dotted line for  the D and B mesons is obtained with the unmodified wave
function \req{wf}, the solid line with  the modified wave function with the scale factor $\al=\half$ in \req{modfac}.}
\end{figure}

In Fig. \ref{DA} we show the distribution amplitudes \req{DA} for
the chiral case and for the heavy pseudoscalar mesons; the dotted
lines for the heavy mesons  correspond to the unmodified
wave function \req{wf}, the solid ones are obtained from the
modified wave function with the scale factor $\al=\half$ in \req{modfac}.

The increasing discrepancy between the  longitudinal
momentum of the light constituents and that of the heavy quark,
with increasing quark mass, could provide a plausible explanation of why
the $\Si_c$ and $\Si_b$ do not fit on the trajectories for a 
pseudoscalar meson.  In this case a scalar diquark
cluster can be formed only by the heavy and a light quark, whereas
the cluster formed of two light quarks has isopin 1 and hence
quark spin 1. The trajectories for the pseudoscalar mesons are
characterized by $s=0$, hence  they are matched to baryons of scalar diquarks. Due
to the increasing difference between the longitudinal momenta, the 
formation of a heavy-light cluster becomes less and less probable
with increasing heavy quark mass. This is also observed:  the mass
difference $\delta_M$  between  the $\Sigma^*_b$, which must
contain a spin 1 cluster, and the $\Sigma_b$  is  $\delta_M=20$
MeV; in contrast,  the $\Si^*_c(2520)$, which must contain a spin 1
cluster, and the $\Si_c(2455)$ is $\de_M= 65$ MeV.



\begin{thebibliography}{99}
\bibitem{notesusy2}
In the present framework, dynamical supersymmetry between mesons
and baryons is not a consequence of supersymmetric QCD at the
level of fundamental fields, but the supersymmetry between the LF
bound-state equations of mesons and baryons.
\end{thebibliography}
\end{document}